\documentclass{webofc}

\usepackage[varg]{txfonts}   
\usepackage{hyperref}
\usepackage{url}
\usepackage{xcolor}
\usepackage{ulem}

\hypersetup{colorlinks=true,citecolor=blue,urlcolor=blue,linkcolor=blue}
\begin{document}
\title{Gamma ray emission from embedded young massive star clusters unveiled by Fermi-LAT}
%
%

\author{\firstname{Giada} \lastname{Peron}\inst{1}\fnsep\thanks{\email{giada.peron@inaf.it}} \and
        \firstname{Giovanni} \lastname{Morlino}\inst{1}\fnsep\and
        \firstname{Elena} \lastname{Amato}\inst{1}\fnsep\and
        \firstname{Stefano} \lastname{Menchiari}\inst{1,2}
}

\institute{INAF Osservatorio Astrofisico di Arcetri, Largo Enrico Fermi,5, 50125, Forence, Italy 
\and Instituto de Astrof\`{ı}sica de Andaluc\`{ı}a, CSIC, 18080 Granada, Spain}

\abstract{Massive star clusters (SCs) have been proposed as additional contributors to Galactic Cosmic rays (CRs), to overcome the limitations of supernova remnants (SNR) to reach the highest energy end of the {Galactic} CR spectrum. Thanks to fast mass losses through collective stellar winds, the environment around SCs is potentially suitable for particle acceleration up to PeV energies. A handful of star clusters has been detected in gamma-rays confirming the idea that particle acceleration is taking place in these environments. Here we present a new analysis of Fermi-LAT data collected towards a few massive young star clusters and estimate the contribution of these types of sources to the bulk of CRs. We then briefly discuss the observational prospects for ASTRI and CTAO.}

%
\maketitle
\section{Introduction}
\label{intro}
Young massive star clusters (YMSCs) are common objects in our Galaxy. Thanks to the {extensive} censuses of stars pursued by Gaia and its precursors, thousands of these objects have been identified, and about as many are hidden behind the gas cocoon (a.k.a. the H\textsc{ii} region) that surrounds them in the earliest phase of their life \cite{Cantat-Gaudin2018AWay,Anderson2014TheRegions}. YMSCs feature very powerful winds, with a total power estimated for the entire Galaxy of the order of 10$^{41}$ erg s$^{-1}$, 
within one order of magnitude of the total power released by Supernove (SNe) in the Milky Way. For this reason they are considered valuable contributors to the Galactic population of CRs, if a non-negligible fraction of this power is converted into accelerated particles. Different acceleration processes could take place in SCs: wind-wind interaction \cite{Reimer2007}, second-order Fermi acceleration through supersonic turbulence induced by winds and SN explosions \citep{Bykov2020High-EnergyRegions, Vieu2022CosmicSuperbubbles, Vink2024}, or particle acceleration at the collective cluster wind termination shock \citep{Morlino2021}. Independent of the acceleration mechanism, if particles are accelerated in these objects, detectable gamma-ray emission is expected. So far, the detection of SCs was limited to a dozen objects in the GeV energy range, while only a handful of objects are seen in the TeV energy range, possibly due to the difficulties in revealing such extended and faint sources by gamma-ray instruments. However, as noted by \cite{Peron2024OnSources}, several of these star clusters could have been detected and labeled as "unidentified" sources. 
The statistical analysis conducted in \cite{Peron2024OnSources} showed that 127 H\textsc{ii} regions of the WISE catalog, hereafter "WISE clusters" \cite{Anderson2012} could have a Fermi-LAT counterpart; {the {randomness} of the association between Fermi and WISE sources was tested with Montecarlo simulations, which assessed that the high degree of superposition between these catalogs could not be due only to to chance coincidence } 
{At the same time, no significant correlation emerged between GeV gamma-ray sources and Gaia-identified SCs (hereafter Gaia clusters \cite{CantatGaudin2020,Celli2024Mass}).}
A fraction of the brightest {Gaia clusters}, though, do have overlapping gamma-ray sources, but to understand whether this is a random coincidence, a deeper analysis is needed. In order to confirm the association and exclude false positives we perform dedicated Fermi-LAT analysis of these objects, which we present in the following for a fraction of the SCs matching gamma-ray sources. {New data collected towards Gaia clusters NGC~3603, NGC~6611, NGC~6618, Berkeley~59 are presented, and compared to the results from WISE clusters RCW~38, RCW~36, and RCW~32 \cite{Peron2024ThePopulation}. }

\section{Gamma-ray observations of embedded clusters}
\begin{figure}
    \centering
    \includegraphics[width=0.245\linewidth, height=0.21\textwidth]{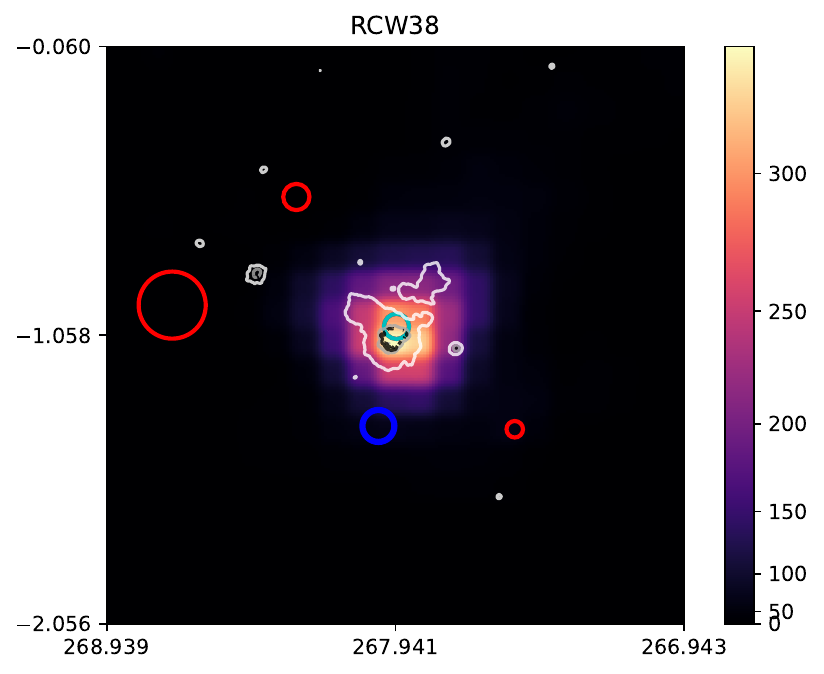}\includegraphics[width=0.245\linewidth, height=0.21 \textwidth]{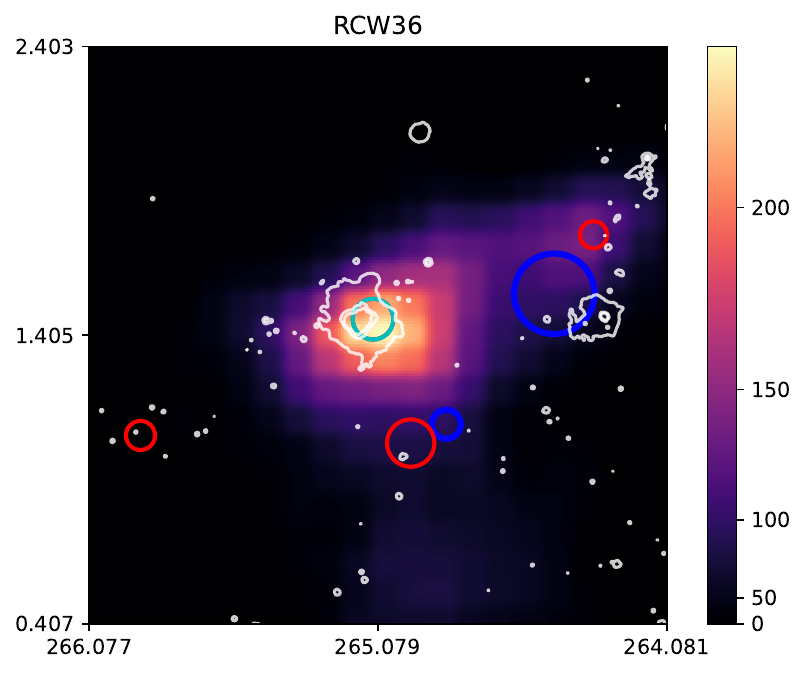}\includegraphics[width=0.245\linewidth, height=0.21 \textwidth]{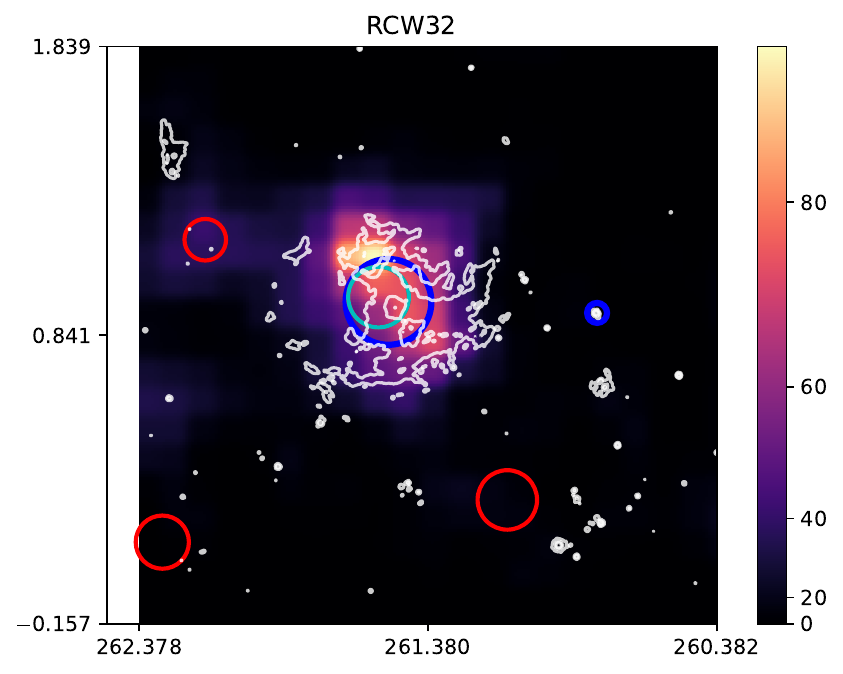}\includegraphics[width=0.245\linewidth, height=0.21 \textwidth]{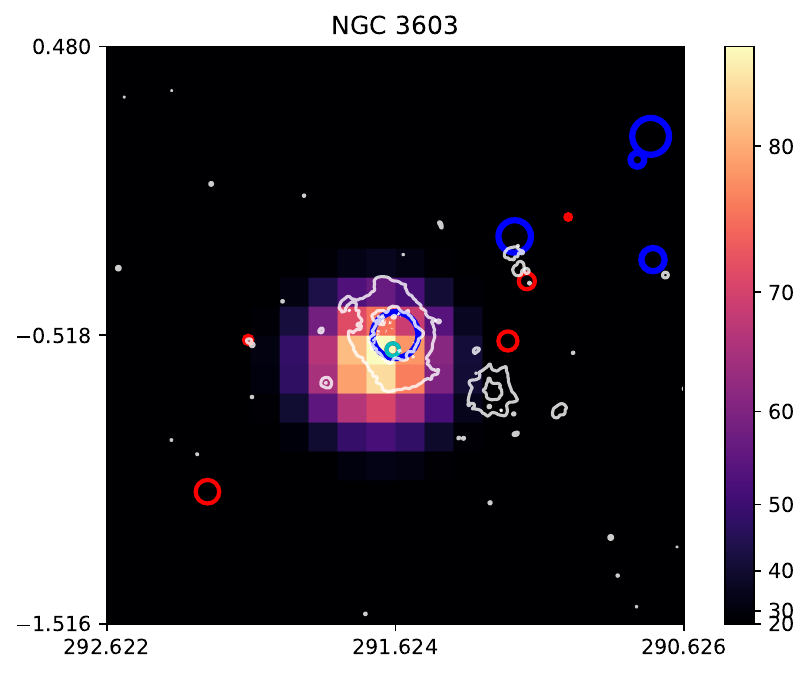}
    
    \includegraphics[width=0.245\linewidth, height=0.21\textwidth]{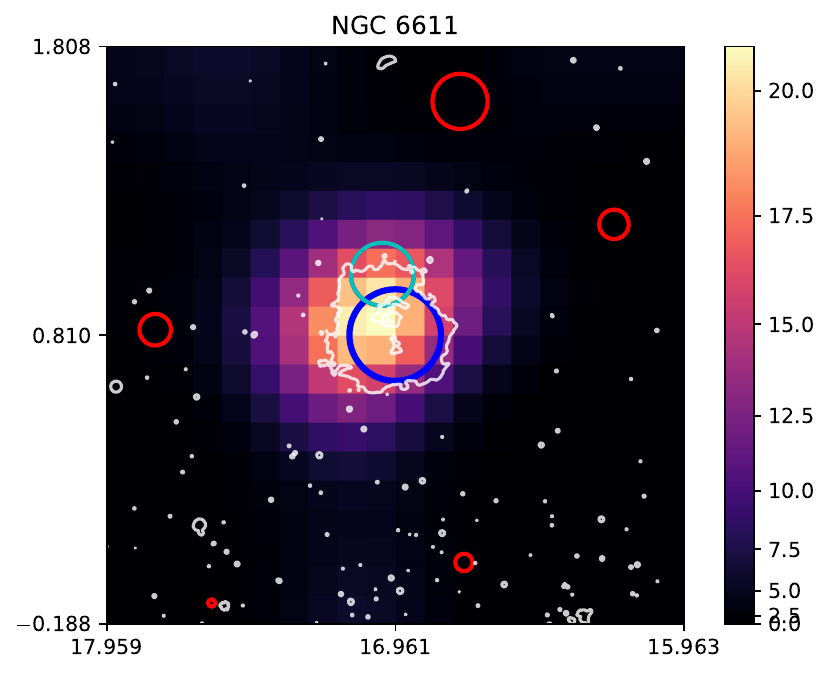}\includegraphics[width=0.245\linewidth, height=0.21 \textwidth]{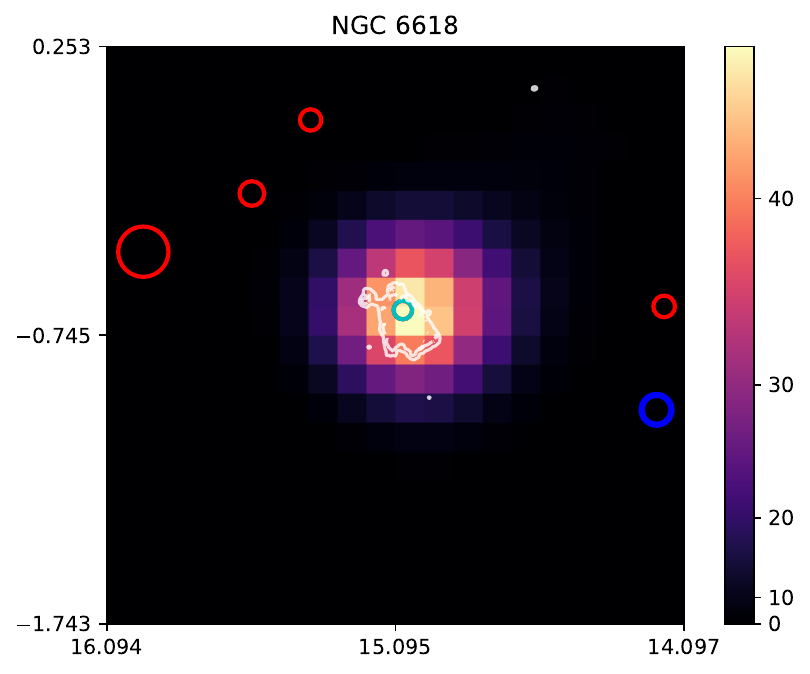}\includegraphics[width=0.245\linewidth, height=0.21 \textwidth]{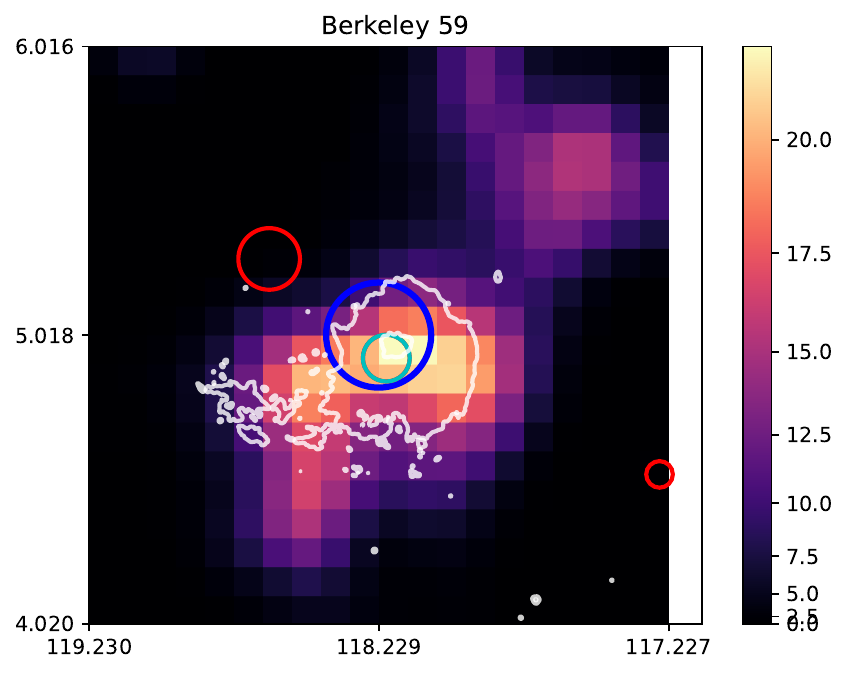}
    \caption{Test statistics maps of the targeted star clusters obtained from Fermi-LAT data. The red circles represent the equivalent radius of the 68\% confidence ellipse of the localization of background sources of the 4FGL catalog \cite{Abdollahi2022IncrementalCatalog}. {The cyan circles represent the confidence radius for the 4FGL sources removed from the background model}.The blue circles are the angular extension of the termination shock for the Gaia star clusters \cite{Celli2024Mass}. The contours trace the 22-$\mu$m emission revealed by WISE in the region \cite{Anderson2012}.}
    \label{fig:tsmaps}
\end{figure}

\begin{figure}
    \centering
    \includegraphics[width=0.5\linewidth]{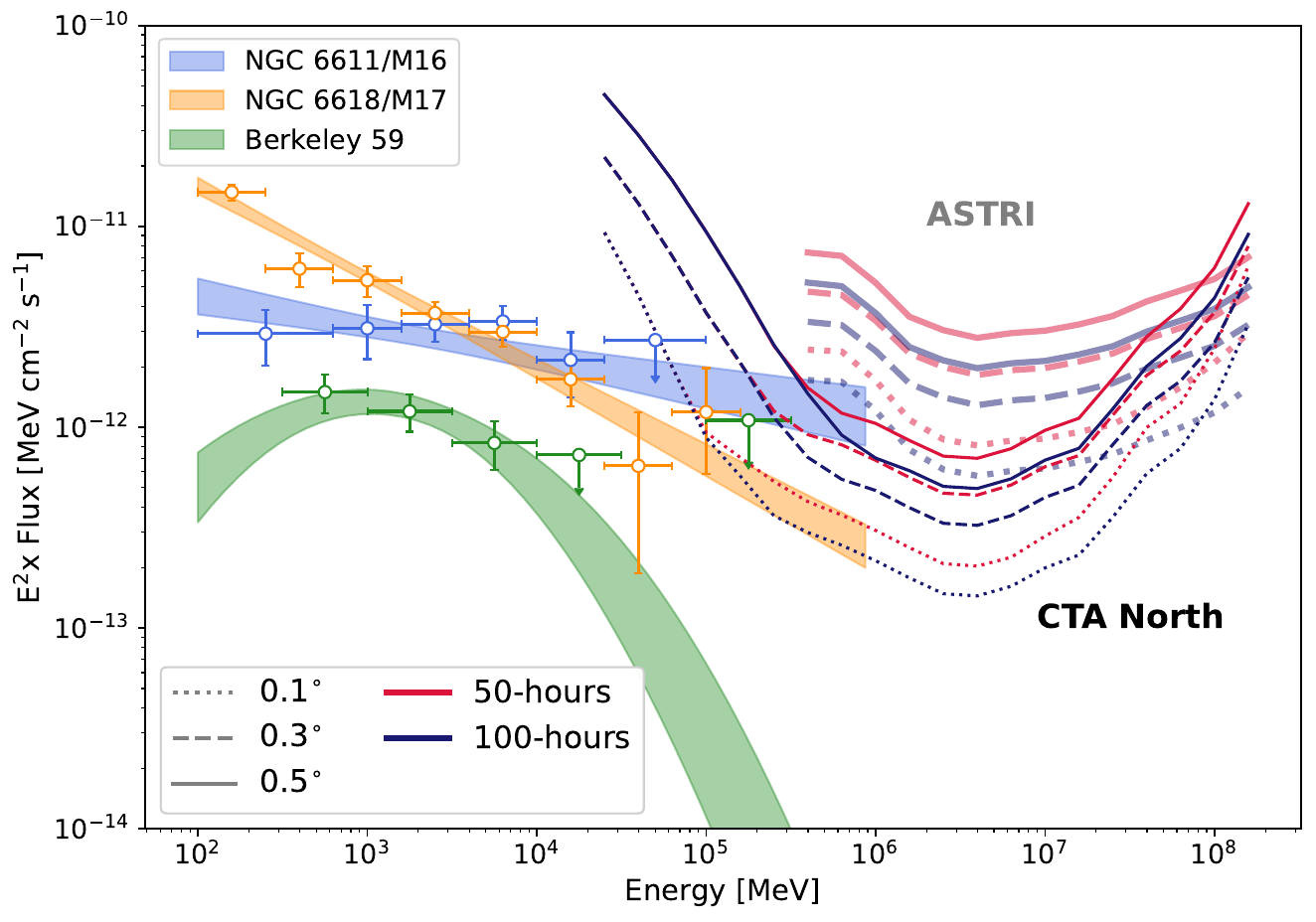}\includegraphics[width=0.5\linewidth]{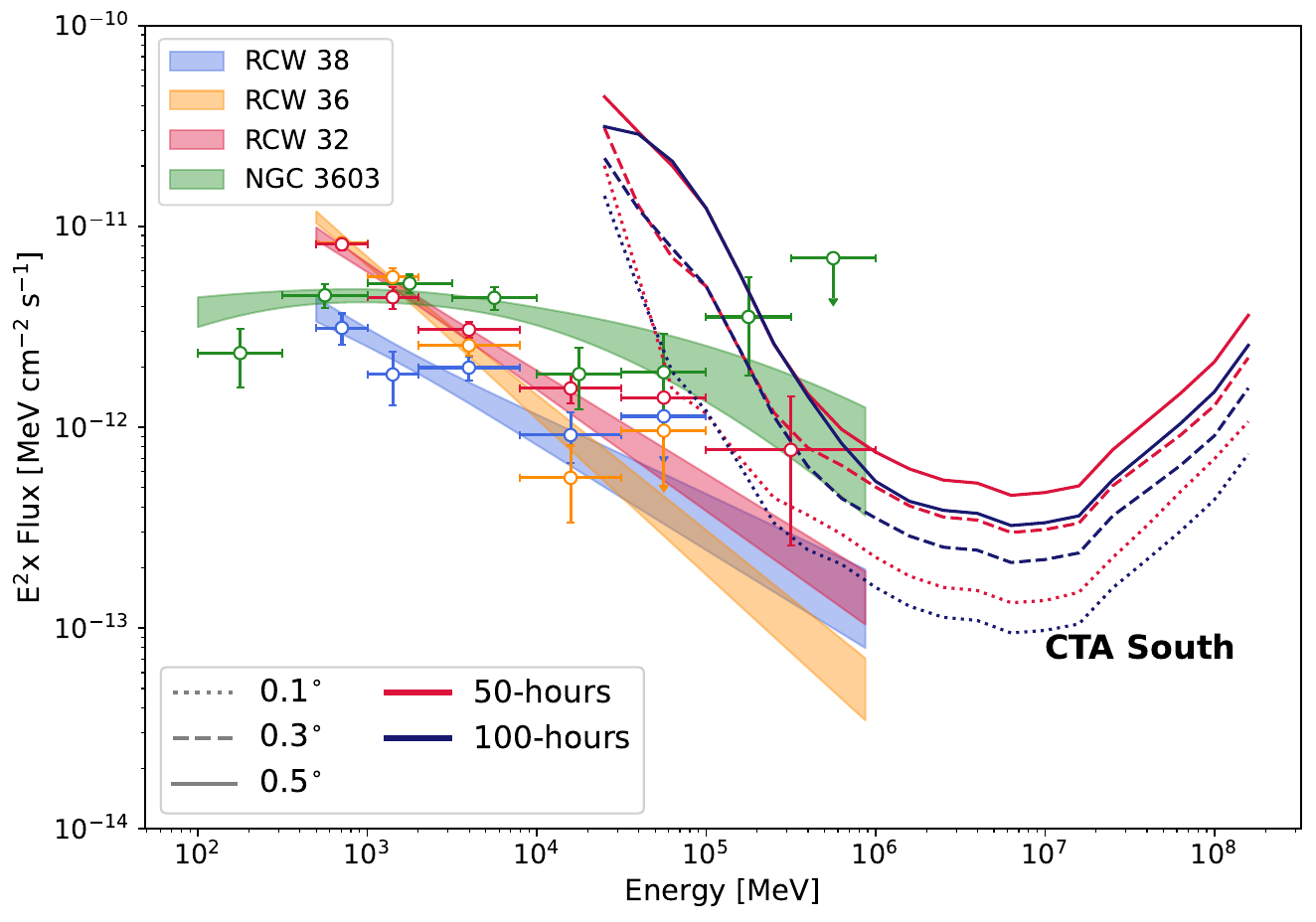}
    \caption{Spectral energy distributions of the targeted SCs, derived from Fermi-LAT analysis, are compared with the sensitivity curves for extended source for the next generation of Cherenkov telescopes, ASTRI and CTAO (the North site on the left, the South site on the right). The curves are obtained following \cite{Celli2024DetectionLHAASO}. On the left side, the SCs that culminate at low zenith angles in the Northern sky are displayed; on the right the ones that culminate at low zenith angles in the Southern sky. }
    \label{fig:vis}
\end{figure}

We perform a standard Fermi-LAT analysis where the background model is composed by the galactic and extragalactic diffuse emission as given by the Fermi-LAT collaboration\footnote{See \href{https://fermi.gsfc.nasa.gov/ssc/data/access/lat/BackgroundModels.html}{https://fermi.gsfc.nasa.gov/ssc/data/access/lat/BackgroundModels.html} {Being all our targeted sources also detected as 4FGL sources, we believe that the choice of the standard background should be safe, nevertheless an extended discussion on the influence of the background choice is reported in \cite{Peron2024ThePopulation} and should in principle be extended to the Gaia SC sample. We defer this to future work.}}, and by the point sources of the 4th Fermi source catalog (4FGL \cite{Abdollahi2022IncrementalCatalog}).  After the optimization of the parameters of the background sources, we remove from the model the sources that are potentially associated with the clusters \cite{Peron2024OnSources}: the resulting residual maps, computed in terms of test statistics (TS) are reported in Figure \ref{fig:tsmaps}. The figure shows the background 4FGL sources in the region as red circles which represent the uncertainty on their localization, as reported in the catalog, while the source associated to the cluster is drawn in cyan. The size of the termination shock for the Gaia star clusters  \cite{Celli2024Mass} is reported as blue circles: according to the model proposed by \cite{Morlino2021}, this should represent the region where particle acceleration takes place.  The contours show the shape of the hot dust as revealed by WISE at 22$\mu$m. Dust grains are heated by the hot radiation produced by the most massive stars of the clusters, and re-emit in the far infrared. The size of the H\textsc{ii} region is in good  approximation comparable with the size of the wind-blown bubble: a thick shell of hot material is indeed expected at the edge of the swept-up region and is often detected in these systems \cite{Anderson2012}. 
Both for WISE and for Gaia clusters targeted clusters an excellent spatial correlation emerges between the Fermi-LAT and the infrared emission, as seen in Figure \ref{fig:tsmaps}. As detailed in \cite{Peron2024ThePopulation}, this suggests an hadronic interpretation for the emission. Large gas densities are indeed expected in bright H\textsc{ii} regions, even of the order of 10$^3$ cm$^{-3}$ and beyond. The spectral energy distribution (SED) derived for each SC is shown in Figure \ref{fig:vis}. As one can see, the spectra appear quite steep, suggesting either that a weak shock is producing the acceleration, or that escape already took place for the high-energy part of the {particle distribution}. Detailed modeling of the acceleration and propagation mechanism is needed to fully describe these objects. However, one can attempt an order of magnitude estimation of the acceleration efficiency that characterizes the targeted SCs. For the WISE clusters, this is done in \cite{Peron2024ThePopulation}, assuming full confinement of the particles, {namely that all particles loose energy before escaping the region, which in formulae reads: $t_{\rm pp}<<t_{\rm esc}$.}
As discussed also in \cite{Peron2024ThePopulation}, this is probably far from real, but {can be used } to derive a stringent lower limit to the acceleration efficiency. Under this assumption we can infer the CR power, $L_{\rm CR}$, from the gamma-ray luminosity, $L_{\gamma}$, as $L_{\rm CR}\approx 3 L_{\gamma}$, where the factor 3 accounts for the three main products of nuclear collisions. 
The resulting values are reported in Table \ref{tab:eff}, where they are compared to the wind luminosity, $L_{w}$, computed by \cite{Celli2024Mass}.
An average value of $\eta_{\min}\equiv L_{\rm CR}/L_{\rm w}$ of about 0.5\% is found for the newly analyzed clusters, compatible to what was found in \cite{Peron2024ThePopulation} . 

\begin{table}[]
    \centering
    \begin{tabular}{|c|ccc|}
\hline
Star Cluster & $L_{w}$ & $L_{\gamma}$ & $\eta_{min}$ \\
\hline
NGC~6611 & 6.3 $\times 10^{36}$ & 7.6$\times 10^{33}$ & 0.36 \% \\
Berkeley~59 & 3.7$\times 10^{36}$ & 6.3$\times 10^{32}$ & 0.05 \% \\
NGC~3603 & 4.6 $\times 10^{37}$& 1.5$\times 10^{35} $ & 0.97 \%\\
\hline
    \end{tabular}
    \caption{Wind and gamma-ray luminosity for the listed clusters. All luminosity are in erg s$^{-1}$. $\eta_{min}$ is the lower limit to the CR acceleration efficiency estimated as described in the text. }
    \label{tab:eff}
\end{table}

\section{Conclusions and prospects}
More and more star clusters are emerging in the gamma-ray band, especially at GeV energies, and observations suggest that they contribute a small but non-negligible fraction of the Galactic CR population. The next steps should aim at constraining this source class at the highest energies. Remarkably, some of the most powerful gamma-ray emitters in the very high-energy (VHE) band, such as Westerlund~1, Westerlund~2 and Cygnus OB2, coincide with SCs, but the small number of cases and the unavoidable source confusion prevent us from obtaining a statistically solid assessment of  their contribution at high energies. A big limitation for detecting SCs is their large extension, that strongly limits the observations of the current high-resolution ground based instruments. {Nevertheless, comparing the derived SEDs with the expected performance for extended sources \cite{Celli2024DetectionLHAASO} of the future Cherenkov observatories, ASTRI and CTAO (Figure \ref{fig:vis}), we expect to be able to target these objects in the next future, taking advantage of both the large field of view and the improved angular resolution, helping us to derive more stringent constraints on the contribution of SCs at VHE.}  

\bibstyle{woc}
\bibliography{bibliography} 
\end{document}